\documentclass[aps,prd,reprint,groupedaddress,twocolumn,showpacs,10pt]{revtex4-1}
\usepackage{graphicx,color}
\usepackage{bm}
\usepackage{amssymb,amsmath}

\newcommand{\be}{\begin{equation}}
\newcommand{\ee}{\end{equation}}
\newcommand{\ba}{\begin{array}}
\newcommand{\ea}{\end{array}}
\newcommand{\bqa}{\begin{eqnarray}}
\newcommand{\eqa}{\end{eqnarray}}

\begin{document}


\title{Comprehending heavy charmonia and their decays by hadron loop effects}


\author{Zhi-Yong Zhou}
\email[]{zhouzhy@seu.edu.cn}
\affiliation{Department of Physics, Southeast University, Nanjing 211189,
P.~R.~China}
\author{Zhiguang Xiao}
\email[]{xiaozg@ustc.edu.cn}
\affiliation{Interdisciplinary Center for Theoretical Study, University of Science
and Technology of China, Hefei, Anhui 230026, China}


\date{\today}

\begin{abstract}
We present that including the hadron loop effects could help us to understand the spectrum of the heavier charmonium-like states and their decays simultaneously.  The observed states could be represented by the poles on the complex energy plane. By coupling to the opened thresholds, the pole positions are shifted from the bare states predicted in the quenched potential model to the complex plane. The pole masses are generally pulled down from the bare masses and the open-charm decay widths are related to the imaginary parts of the pole positions. Moreover, we also analyze the pole trajectory of the $\chi_{c1}(2P)$ state while the quark pair production rate from the vacuum changes in its uncertainty region, which indicates that the enigmatic $X(3872)$ state may be regarded as a $1^{++}$ $c\bar{c}$ charmonium-dominated state dressed by the hadron loops as the others.
\end{abstract}


\maketitle
\section{Introduction}

Before the $X(3872)$ was found~\cite{Choi:2003ue}, there were only four
well-established charmonium states above the $D\bar{D}$ threshold. In
recent years, along with explosion of the experimental activities
on the heavy quarkonium physics, more than a dozen of charmonium-like
$``XYZ"$ states above the open-flavor thresholds have been observed
and the charmonium family is remarkably enriched. Until now, there are
fourteen neutral charmonium-like states quoted in the Particle Data
Group~(PDG) Table~\cite{Beringer:1900zz}. However, the masses of most
newly observed states above open-charm thresholds run out of the
predictions of the quark potential model~\cite{Godfrey:1985xj} which
proved to be successful for those states below the $D\bar{D}$
threshold. Therefore, different approaches are adopted to understand
them case by case and there is no consensus on the natures of those
``unexpected" states.

$X(3872)$ is a typical example in this situation. Its mass is too low
to be a $2P$ $c\bar{c}$ state in the potential
model~\cite{Barnes:2003vb} and this possibility was almost given up,
after the isospin violating decay $X\rightarrow J/\psi\rho$ was
confirmed. As the state is located just at the $D\bar{D}^*$ threshold,
it is also suggested to be a $D\bar{D}^*$ molecule bounded by pion
exchanges~\cite{Tornqvist:2004qy,Close:2003sg,Voloshin:2003nt,Wong:2003xk,Braaten:2003he}.
This assignment can explain the properties of the mass and the
$J^{PC}$ of $X(3872)$, but it encounters serious problems in other
aspects. For example, as a loosely bounded $D\bar{D}^*$ molecule, it
is difficult to radiatively transit into excited charmonium states,
such as $\psi'$, through the quark annihilation or other mechanisms.
The BaBar collaboration~\cite{Aubert:2008ae} measured the ratio $
{Br(X\rightarrow \psi'\gamma)/ Br(X\rightarrow \psi\gamma)}=3.4\pm
1.4, $ which is several orders of magnitude higher than the model
predictions, e.g. in Ref.~\cite{Swanson:2003tb,Swanson:2004pp}.

The difficulties also remind theorists that the vacuum fluctuation
effect should receive more attention in understanding the heavier
charmonia. In the quark potential models, a charmonium state is
considered as a bound state of a charm quark and its antiquark through
a non-relativistic interaction potential, typically incorporating a
Coulomb term at a short distance and a linear confining term at a
large distance. These models neglect the modifications due to quantum
fluctuations, i.e., the creation of light quark pairs, which can be
represented by the hadron loops in the coupled
channel model.  This coupled channel effect was considered in the
Cornell model,~\cite{Eichten:1978tg} and it  has  also been used to
study the resonances with strongly coupled S-wave thresholds, where
the states are drawn to their strongly coupled
thresholds~\cite{vanBeveren:1983td}.  In particular,
Heikkil$\ddot{\mathrm{a}}$, T$\ddot{\mathrm{o}}$rnqvist, and Ono
developed a unitarized quark model,  carrying over the Dyson summation
idea, to study the charmonium spectrum long
ago~\cite{Heikkila:1983wd}. Recently, Pennington and
Wilson~\cite{Pennington:2007xr} extracted the mass shifts of
charmonium states from the results of a non-relativistic potential
model by Barnes, Godfrey, and Swanson~\cite{Barnes:2005pb} by
considering the hadron loop effect. K.T.Chao and his collaborators
also proposed a screened potential model~\cite{Chao:1989ek,Li:2009zu} to
investigate the heavy quarkonium spectrum, in which the effect of
vacuum polarization is incorporated in a different way.

Our present study goes along the same lines as Ref.~\cite{Pennington:2007xr} with several significant improvements. First,
instead of using an empirical universal form factor to describe the
coupling vertices between the charmonium states and the decaying
channels as in~\cite{Pennington:2007xr}, we formulate the vertex
functions by adopting the $^3P_0$ model so that they could be
represented by the parameters in the potential model.  Secondly,
incorporating the $^3P_0$ model into this scheme also enables us to
produce not only the mass shifts but also the decay widths, whereas
the decay widths are inputs in Ref.~\cite{Pennington:2007xr} extracted
from Ref.~\cite{Barnes:2005pb}. Moreover, this analytical formulation
also shows more merits by allowing us to explore the poles on the
complex energy plane, whose behaviors as the parameters change shed
more insight on the nature of these states, especially with regard to
the enigmatic $X(3872)$ state. It is also worth mentioning that this
calculation covers all the related charmonium states in one unified
picture instead of treating them case by case.

In this study, we found that the discrepancies between the observed
masses  and the predictions of the quenched potential model could be
compensated by taking the hadron loop effect into account. Meanwhile,
their open-charm decay widths are reproduced in a reasonable manner.
That means, most of the states discussed in this paper could be
depicted in a unified picture, as a charmonium state dressed by hadron
loops, or similarly, as a mixture of a conventional charmonium state
and the coupled continuums. The enigmatic $X(3872)$ could also be
included in this scheme without any ``exotic" aspect.

The paper is organized as follows: In Section II, the main scheme and
how to model the coupled channels are briefly introduced. Numerical
procedures and results are discussed in Section III. Section IV is
devoted to our conclusions and further discussions.

\section{The model}
In a non-relativistic quark potential model, a quarkonium meson is
regarded as a bound state of a quark and an anti-quark, formed by the
effective potential generated from the gluon exchange diagrams and, in
certain circumstances, the annihilation diagrams. At the hadron level,
the bare propagator of such a bound state could be represented as
\bqa\mathbb{P}(s)=1/(m_0^2-s),\eqa
with a pole on the real axis of the complex $s$ plane, corresponding
to a non-decaying state, where $m_0$ is the mass of the ``bare"
$q\bar{q}$ state. Once its coupling to certain two-body channels is
considered, the inverse meson propagator, $\mathbb{P}^{-1}(s)$, is
expressed as
\bqa\mathbb{P}^{-1}(s)=m_0^2-s+\Pi(s)=m_0^2-s+\sum_n\Pi_n(s),\eqa  and
$\Pi_n(s)$ is the self-energy function for the $n$-th coupling
channel. Here, the sum is over all the opened channels and, in
principle, all virtual channels.  $\Pi_n(s)$ is an analytic function
with only a right-hand cut starting from the $n$-th threshold
$s_{th,n}$, and so, one can write down its real part from its
imaginary part through a dispersion relation
\bqa\label{dr}
\mathrm{Re}\Pi_n(s)=\frac{1}{\pi}\mathcal{P}\int_{s_{th,n}}^\infty\mathrm{d}z\frac{\mathrm{Im}\Pi_n(z)}{(z-s)},
\eqa
where $\mathcal{P}\int$ means the principal value integration.  The
mass and total width of a meson are specified by a pole of
$\mathbb{P}(s)$ on the unphysical Riemann sheet attached to the
physical region, usually defined as
$s_{pole}=(M_{pole}-i\Gamma_{pole}/2)^2$.  This mechanism is typified
by the Dyson-Schwinger equation for the propagator of the $\rho$ meson
as illustrated in Ref.\cite{Pennington:2010dc}. As the bare $\rho$
state predominantly couples to the $\pi\pi$ system in the $P$ wave,
the pole will move away from the real axis onto the complex energy
plane, thus the $\rho$ meson could be regarded as largely a $q\bar{q}$
state with a few percent $\pi\pi$.~\cite{Geiger:1992va}

To investigate the pole positions in this scheme, we make use of the
Quark Pair Creation~(QPC)
model~\cite{Micu:1968mk,Colglazier,LeYaouanc:1972ae}, also known as
the $^3P_0$ model in the literature, to model the coupling vertices of
the imaginary part of the self-energy function in this calculation.
This is not only because this model has proved to be successful in
many phenomenological calculations but also because it could provide
analytical expressions of the vertex functions.  Furthermore, the
exponential factors in the vertex functions of the QPC model provide a
natural ultraviolet suppression to the dispersion relation, which is
chosen by hand according to the empirical strong interaction length
scale in Ref.~\cite{Pennington:2007xr}.

A modern review of the QPC model and calculation of the transition
amplitude can be found in Ref.\cite{Luo:2009wu}. The main ingredients
of this model are summarized in the following. In the QPC model, a
meson~(with a quark $q_1$ and an anti-quark $q_2$) decay occurs by
producing a quark~($q_3$) and anti-quark~($q_4$) pair from the vacuum.
In the non-relativistic limit, the transition operator is represented
as
\bqa
T=-3\gamma\sum_m\langle 1 m 1 -m|00\rangle\int d^3\vec{p_3}d^3\vec{p_4}\delta^3(\vec{p_3}+\vec{p_4})\nonumber\\ \mathcal{Y}_1^m(\frac{\vec{p_3}-\vec{p_4}}{2})\chi_{1 -m}^{34}\phi_0^{34}\omega_0^{34}b_3^\dagger(\vec{p_3})d_4^\dagger(\vec{p_4}),
\eqa
where $\gamma$ is a dimensionless parameter to represent the quark pair production rate from the vacuum, and
$\mathcal{Y}_1^m(\vec{p})\equiv p^lY_l^m(\theta_p,\phi_p)$ is a solid
harmonic function that gives the momentum-space distribution of the created
pair. Here the spins and relative orbital angular momentum of the
created quark and anti-quark~(referred to by subscripts $3$ and $4$,
respectively) are combined to give the overall
$J^{PC}=0^{++}$ quantum numbers.
$\phi_0^{34}=(u\bar{u}+d\bar{d}+s\bar{s})/\sqrt{3}$ and
$\omega_0^{34}=\delta_{ij}$, where $i$ and $j$ are the SU(3)-color
indices of the created quark and anti-quark. $\chi_{1 -m}^{34}$ is a
triplet of spin.
The helicity amplitude $\mathcal{M}^{M_{J_A},M_{J_B},M_{J_C}}$ is from the transition amplitude
\bqa
\langle BC|T|A \rangle =\delta^3(K_B+K_C-K_A)\mathcal{M}^{M_{J_A},M_{J_B},M_{J_C}}.
\eqa
(see Ref.\cite{Luo:2009wu} for the details).

Thus, the imaginary part of the self-energy function in the dispersion
relation, Eq.(\ref{dr}), could be expressed as
\bqa\label{discont}
Im\Pi_{A\rightarrow BC}(s)=-\frac{\pi^2}{2J_A+1} \frac{|\vec{P}(s)|}{\sqrt{s}}\nonumber\\
\sum_{M_{J_A},M_{J_B},M_{J_C}}|\mathcal{M}^{M_{J_A},M_{J_B},M_{J_C}}(s)|^2,
\eqa
where $|P(s)|$ is the three-momentum of $B$ and $C$ in their center of mass frame.  So,
\bqa
\frac{|P(s)|}{\sqrt{s}}=\frac{\sqrt{(s-(m_B+m_C)^2)(s-(m_B-m_C)^2)}}{2s}.
\eqa

The $A-BC$ amplitude reads

\begin{widetext}
\bqa
\mathcal{M}^{M_{J_A}M_{J_B}M_{J_C}}(\vec{P})&=&\gamma\sqrt{8E_AE_BE_C}\sum_{M_{L_A},M_{S_A},M_{L_B},M_{S_B},M_{L_C},M_{S_C},m}\langle L_A M_{L_A}S_A M_{S_A}|J_A M_{J_A}\rangle\nonumber\\
&&\times\langle L_B M_{L_B}S_B M_{S_B}|J_B M_{J_B}\rangle\langle L_C M_{L_C}S_C M_{S_C}|J_C M_{J_C}\rangle \langle 1 m 1 -m|00\rangle\nonumber\\
&&\times \langle \chi_{S_C M_{S_C}}^{32}\chi_{S_B M_{S_B}}^{14}|\chi_{S_A M_{S_A}}^{12}\chi_{1 -m}^{34}\rangle \langle \phi_C^{32}\phi_B^{14}|\phi_A^{12}\phi_0^{34}\rangle I_{M_{L_B},M_{L_C}}^{M_{L_A},m}(\vec{P}).
\eqa
\end{widetext}
The spatial integral $I_{M_{L_B},M_{L_C}}^{M_{L_A},m}(\vec{P})$ is given by
\bqa
& I_{M_{L_B},M_{L_C}}^{M_{L_A},m}(\vec{P})=\int d^3\vec{k}\psi^*_{n_B
L_B M_{L_B}}(-\vec{k}+\frac{\mu_4}{\mu_1+\mu_4}\vec{P})\nonumber\\ & \times\psi^*_{n_C L_C
M_{L_C}}(\vec{k}-\frac{\mu_3}{\mu_2+\mu_3}\vec{P})\psi_{n_A L_A
M_{L_A}}(-\vec{k}+\vec{P})\mathcal{Y}_1^m(\vec{k}),\nonumber\\
\eqa
where we have taken $\vec{P}\equiv \vec{P_B}=-\vec{P_C}$ and $\mu_i$
is the mass of the $i$-th quark. $\psi_{n_A L_A M_{L_A}}(\vec{k_A})$
is the relative wave function of the quarks in meson $A$ in the
momentum space.

The recoupling of the spin  matrix element can be written, in terms of
the Wigner's $9$-$j$ symbol, as
\bqa
&&\langle \chi_{S_C M_{S_C}}^{32}\chi_{S_B M_{S_B}}^{14}|\chi_{S_A M_{S_A}}^{12}\chi_{1 -m}^{34}\rangle
\nonumber\\ &&=[3(2S_B+1)(2S_C+1)(2S_A+1)]^{1/2}\nonumber\\ &&\times\sum_{S,M_S}\langle S_C M_{S_C}S_B M_{S_B}|S M_S\rangle \langle S M_S|S_A M_{S_A};1,-m\rangle
\nonumber\\ &&\times\left\{
\begin{array}{ccc}
                                                                                                1/2 & 1/2 & S_C \\
                                                                                                1/2 & 1/2 & S_B \\
                                                                                                S_A & 1 & S \\
                                                                                          \end{array}
											  \right\}.
\eqa
The flavor matrix element is
\bqa
&&\langle\phi_C^{32}\phi_B^{14}|\phi_A^{12}\phi_0^{34}\rangle\nonumber\\&&=\sum_{I,I^3}\langle I_C,I_C^3;I_B I_B^3|I_A I_A^3\rangle
\nonumber\\ &&\times[(2I_B+1)(2I_C+1)(2I_A+1)]^{1/2}\left\{\begin{array}{ccc}
                                                                                                I_2 & I_3 & I_C \\
                                                                                                I_1 & I_4 & I_B \\
                                                                                                I_A & 0 & I_A \\
                                                                                              \end{array}
                                                                                            \right\},\nonumber\\
\eqa
where $I_i (I_1, I_2, I_3, I_4)$ is the isospin of the quark $q_i$.

With all the necessary ingredients of the model at hand, care must be taken when Eq.(\ref{discont}) is continued to the complex
$s$ plane for extracting the related poles. Since what is used in this model is only the tree level
amplitude, there is no right hand cut for the amplitude,
$\mathcal{M}^{M_{J_A},M_{J_B},M_{J_C}}(s)$. Thus, the analytical
continuation of the amplitude obeys
$\mathcal{M}(s+i\epsilon)^*=\mathcal{M}(s-i\epsilon)=\mathcal{M}(s+i\epsilon)$.
The physical amplitude, incorporating the loop contributions, should have
right hand cuts, and, in principle, the analytical continuation turns
to be
$\mathcal{M}(s+i\epsilon)^*=\mathcal{M}(s-i\epsilon)=\mathcal{M}^{II}(s+i\epsilon)$
by meeting the need of real analyticity.
$\mathcal{M}^{II}(s+i\epsilon)$ means the amplitude on the unphysical
Riemann sheet attached with the physical region.

The general character of the poles on different Riemann sheets has
been discussed widely in the literature, (see, for
example,~\cite{Eden:1964zz}).  A resonance is represented by a pair of conjugate poles
on the Riemann sheet, as required by the real analyticity. The micro-causality
tells us the first Riemann sheet is free of complex-valued poles, and
the resonances are represented by those poles on unphysical sheets.
The resonance behavior is only significantly influenced by those
nearby poles, and that is why only those closest poles to the
experiment region could be extracted from the experiment data in
a phenomenological study.  Those poles on the other sheets, which are
reached indirectly, make less contribution and are thus harder to
determine.

\section{Numerical calculations and discussions}

In this scheme, all the intermediate hadron loops will contribute to
the ``renormalization" of the ``bare" mass of a bound state. It is
easy to find that an opened channel will contribute both a real part
and an imaginary part to the self-energy function, but a virtual
channel contributes only a real one. To avoid counting the
contributions of all virtual channels, we adopt a once-subtracted
dispersion relation, as proposed in Ref.~\cite{Pennington:2007xr}, to
suppress contributions of the faraway ``virtual" channels and to make the picture simpler.  It is reasonable to expect that this lowest charmonium state, as a deep bound state, has the mass defined by the
potential model, uninfluenced by the effect of the hadron loops. Its
mass then essentially defines the mass scale and fixes the
subtraction point. Thus, the subtraction point $s_0$ is chosen at the mass square of the $J/\psi$ state. The inverse of the meson
propagator turns out to be
\bqa
\mathbb{P}^{-1}(s)=m_{pot}^2-s+\sum_n\frac{s-s_0}{\pi}\int_{s_{th,n}}^\infty\mathrm{d}z\frac{\mathrm{Im}\Pi_n(z)}{(z-s_0)(z-s)},\nonumber\\
\eqa
where $m_{pot}$ is the bare mass of a certain meson defined in the potential model.

The bare masses of the charmonium-like states in this calculation are chosen at the values
of the classic work by Godfrey and Isgur~(Refered to GI in the
following)~\cite{Godfrey:1985xj}. The reason why we choose this set of
values is that they provide globally reasonable predictions to meson
spectra with $u$, $d$, $s$, $c$, and $b$ quarks, especially those states below open-flavor thresholds. Thus, the constants
used in our calculation of the coupling vertex is defined as the
values in Ref.~\cite{Godfrey:1985xj} for consistency.  The constituent quark masses are $M_c=1.628\mathrm{GeV}$,
$M_s=0.419\mathrm{GeV}$, and $M_u=0.22\mathrm{GeV}$. The physical masses concerned in the final states are the average values
in the PDG table.~\cite{Beringer:1900zz}
We use the Simple Harmonic Oscillator~(SHO) wave function to represent the relative wave function of quarks in a meson, as usually used in the QPC model calculation. The SHO wave function scale, denoted as the $\beta$ parameter, is from Ref.\cite{Godfrey:1986wj,Kokoski:1985is}, which is chosen to reproduce the root mean square radius of the quark model state. For $D^0$, $D^\pm$, $D^{*0}$, $D^{*\pm}$, $D_s^\pm$, $D_s^{*\pm}$, the $\beta$ values are $0.66$, $0.66 $, $0.54 $, $0.54 $, $0.71 $, $0.59 $, respectively, with units of $\mathrm{GeV}$. The $\beta$ parameters of the charmonium states of the GI's model are reasonably estimated at a universal value of 0.4 $\mathrm{GeV}$, which is consistent with the average value in Ref.~\cite{Close:2005se}. The dimensionless strength parameter is chosen at $\gamma=6.9$ for non-strange $q\bar{q}$ production, and $\gamma_s=\gamma/\sqrt{3}$ for $s\bar{s}$ production.\cite{yaouancbook}

It is also reasonable to assume that the parameters in the GI's work have
included part of the coupled channel effects, especially those virtual hadron loops, because part of the spectrum with open flavor thresholds have been covered in their fit. Furthermore, their predictions to those states below the $D\bar{D}$ threshold are quite precise, which also means the ``renormalization" effects of those virtual hadron loops, which only contribute  real parts in the dispersion relation, have entered the parameters. Thus, to avoid double counting, only those channels that could be open for a certain state are considered, as shown in Table~\ref{channels}.

In a general view, for the $3S$, $4S$, $2P$, $3P$, $1D$, and $2D$
states discussed in this paper, the pole masses are lower by about
10-110$\mathrm{MeV}$ than the predicted values in the quenched
potential model, and the obtained pole masses  agree better with the
measured values. Moreover, even though the $^3P_0$ coupling is
embedded in the scheme and the model parameters are fixed, the extracted pole widths are still reasonable compared with the experimental values for most of the states.

\begin{table*}
\caption{\label{channels}The coupled channels of different states considered in this paper. The $D^0\bar{D}^0$ and $D^+\bar{D}^-$ modes are included in  ``$D\bar{D}$" and it is similar for ``$D\bar{D}^*$" and ``$D^*\bar{D}^*$". The inclusion of the charge conjugate mode is always implied in this paper.
}
\begin{center}
\begin{tabular}{|c|cccccc|}
  \hline
  State ($n^{2S+1}L_J$) & $\ \ \ D\bar{D}\ \ \ $ & $\ \ \ D\bar{D}^*\ \ \ $& $\ D^*\bar{D}^*\ $ & $\ \ \ D_s^+ D_s^-$ & $D_s^+D_s^{*-}$ & $D_s^{*+}D_s^{*-}$  \\
  \hline

  \hline
   $\psi(3^3S_1)$  &$\Box$   &$\Box$  &$\Box$  &$\Box$  &  &  \\
   $\eta_c(3^1S_0)$  &  &$\Box$  &  &   &   &  \\

\hline
  $\psi(4^3S_1)$  &$\Box$  &$\Box$  &$\Box$  &$\Box$  &$\Box$  &$\Box$ \\
 $\eta_c(4^1S_0)$  &   &$\Box$  &$\Box$  &  &$\Box$  &$\Box$ \\

\hline
  $\chi_2(2^3P_2)$  &$\Box$  &$\Box$  &   &$\Box$  &   &  \\
   $\chi_1(2^3P_1)$  &   &$\Box$  &   &  &  &  \\
   $\chi_0(2^3P_0)$  &$\Box$  &  &   &   &   &  \\
    $h_c(2^1P_1)$  &  &$\Box$  &   &   &   & \\

\hline
   $\chi_2(3^3P_2)$ &$\Box$  &$\Box$  &$\Box$  &$\Box$  &$\Box$  &$\Box$ \\
   $\chi_1(3^3P_1)$ &  &$\Box$  &$\Box$  &  &$\Box$  &$\Box$ \\
   $\chi_0(3^3P_0)$ &$\Box$  &   &$\Box$  &$\Box$  &  &$\Box$ \\
     $h_c(3^1P_1)$  &   &$\Box$  &$\Box$  &   &$\Box$  &$\Box$ \\

\hline
  $\psi_3(1^3D_3)$ &$\Box$  &  &   &  &   &  \\
   $\psi(1^3D_1)$  &$\Box$  &   &   &   &   &  \\

\hline
   $\psi_3(2^3D_3)$ &$\Box$  &$\Box$  &$\Box$  &$\Box$  &$\Box$  &$\Box$ \\
   $\psi_2(2^3D_2)$  &  &$\Box$  &$\Box$  &  &$\Box$  &$\Box$ \\
    $\psi(2^3D_1)$  &$\Box$  &$\Box$  &$\Box$  &$\Box$  &$\Box$  &  \\
    $\eta_{c2}(2^1D_2)$  &   &$\Box$  &$\Box$  &   &$\Box$  &  \\

  \hline
  \end{tabular}
\end{center}
\end{table*}

\begin{table*}
\caption{\label{charmlike}The compilation of the pole masses ($\mathrm{Re}[\sqrt{s_{pole}}]$) and the pole widths (2$\mathrm{Im}[\sqrt{s_{pole}}]$) of the states shifted by the hadron loops effects, compared to the observed values and the GI's values. The unit is $\mathrm{MeV}$.
}
\begin{center}
\begin{tabular}{|c|c|c|c|c|c|c|c|c|}
  \hline
  Multiplet & State ($n^{2S+1}L_J$) & $J^{PC}$ & PDG State & Expt. mass & Expt. width & $\mathrm{Re}[\sqrt{s_{pole}}]$ & 2$\mathrm{Im}[\sqrt{s_{pole}}]$  & GI mass \\
  \hline

  \hline
  3S  & $\psi(3^3S_1)$  & $1^{--}$&$\psi(4040)$ & 4039$\pm$1 & 80$\pm$10  & 4051 &  25    & 4100\\
      & $\eta_c(3^1S_0)$ & $0^{-+}$ &  &    &   & 4025  &  23    & 4064\\

\hline
  4S  & $\psi(4^3S_1)$ & $1^{--}$ & $X(4360)$ & 4361$\pm$13 &  74$\pm$18  & 4371 &  49   & 4450\\
      & $\eta_c(4^1S_0)$ & $0^{-+}$ &  &   &   & 4348 &  48    & 4425\\

\hline
  2P  & $\chi_2(2^3P_2)$ & $2^{++}$& $\chi_{c2}'$ & 3927$\pm$2  & 24$\pm$6  & 3942 &  2    & 3979\\
      & $\chi_1(2^3P_1)$ & $1^{++}$ & $X(3872)$ & 3872     & $<$2.3 & 3884 &   4   & 3953\\
      & $\chi_0(2^3P_0)$ & $0^{++}$ & $X(3880)$? & 3878$\pm$48?    & 347$^{+316}_{-143}$?   & 3814 &  133    & 3916\\
      & $h_c(2^1P_1)$ & $1^{+-}$ & $X(3940)$ & 3942$\pm$9    & 37$^{+27}_{-17}$   & 3900 &   6  & 3956\\

\hline
  3P  & $\chi_2(3^3P_2)$ & $2^{++}$ &  &  &   & 4244 &  24    & 4337\\
      & $\chi_1(3^3P_1)$ & $1^{++}$ &   &    &    & 4217 &  84   & 4317\\
      & $\chi_0(3^3P_0)$ & $0^{++}$ &$X(4160)$ & 4160$^{+29}_{-25}$ & 139$^{+110}_{-60}$    & 4210 &  114  & 4292\\
      & $h_c(3^1P_1)$ & $1^{+-}$ &   &    &    & 4219 &  49   & 4318\\

\hline
  1D  & $\psi_3(1^3D_3)$& $3^{--}$ & &   &    & 3838 &  1   & 3849\\
      & $\psi_2(1^3D_2)$& $2^{--}$  &  &     &    &  &     & 3838\\
      & $\psi(1^3D_1)$ & $1^{--}$ & $\psi(3770)$ &  3773   &  27$\pm$1  & 3764 &  18   & 3819\\
      & $\eta_{c2}(1^1D_2)$& $2^{-+}$  &    &   &    &  &     & 3837\\

\hline
  2D  & $\psi_3(2^3D_3)$& $3^{--}$ & &   &   & 4113 & 6    & 4217\\
      & $\psi_2(2^3D_2)$ & $2^{--}$ &   &    &    & 4141 & 72    & 4208\\
      & $\psi(2^3D_1)$ & $1^{--}$ & $\psi(4160)$ &  4153$\pm$3   & 103$\pm$8   & 4080 &   114  & 4194\\
      & $\eta_{c2}(2^1D_2)$ & $2^{-+}$ &   &    &   & 4101 &   44   & 4208\\

  \hline
  \end{tabular}
\end{center}
\end{table*}

The potential model mass of the $\psi(1^3D_1)$ state is at 3819
$\mathrm{MeV}$, while its pole is shifted down to
$\sqrt{s}=$$(3.765-0.009 i)$ $\mathrm{GeV}$ mass, which means that the
pole mass is 3765 $\mathrm{MeV}$ and the pole width is 18
$\mathrm{MeV}$. These values are compatible with that of the observed
$\psi(3770)$ state. The mass and width of $\psi(3770)$ are usually
used to fit the model parameters, so this compatibility  demonstrates the reasonability of the parameters we choose.  Here, the $2S-1D$ mixing is not considered, since the mass of $\psi(2^3S_1)$ predicted by the GI's model is 3680 $\mathrm{MeV}$, which is below the $D^0\bar{D}^0$ threshold that it has no common opened  channels with the $\psi(1^3D_1)$. The mixing through their common virtual channels, which might leads to the large leptonic width of $X(3770)$~\cite{Rosner:2004wy}, is beyond the scope of this study.

Since the masses of all the states are generally pulled down by
considering the hadron loop effect and the spectrum is compressed, the
$\psi(4415)$ seems to be too high to be assigned as the
$\psi(4^3S_1)$, although this assignment is held by the quenched
potential model calculation~\cite{Godfrey:1985xj}. The screened
potential model~\cite{Li:2009zu}, which takes the vacuum fluctuation
into account by introducing a screened potential, also suggests a
compressed spectrum, in which the $\psi(4415)$ is proposed to have a $\psi(5^3S_1)$ assignment. The pole properties of the $\psi(4^3S_1)$ are more compatible with the $X(4360)$ state in this calculation.

In this paper, we did not find the space to accommodate the vector $X(4260)$ state,
which is discovered by the BaBar Collaboration~\cite{Aubert:2005rm}. Actually, there
have already existed some difficulties to assign $X(4260)$ as a
conventional charmonium in the liturature. The most serious one seems to be the observed dip rather than a peak
in the $R$ value scanned in $e^+e^-$ annihilation~\cite{Beringer:1900zz} around
$X(4260)$. Thus, this state could totally or partly be interpreted as
an exotic state, such as a $c\bar{c}g$ hybrid~\cite{Zhu:2005hp}, a
tetro-quark
state~\cite{Maiani:2005pe,Ebert:2008kb,Albuquerque:2008up}, a bayonium
state~\cite{Qiao:2007ce},  or a molecule state~\cite{Ding:2008gr}.  Another possibility for the dip is the destructive interference
among the nearby resonances~\cite{Chen:2010nv}.

The $X(3872)$ was first observed by Belle~\cite{Choi:2003ue} in the $J/\psi\pi^+\pi^-$ invariant mass distribution in $B^+\rightarrow K^+J/\psi\pi^+\pi^-$ decay as a very narrow peak around 3872 $\mathrm{MeV}$ with a width smaller than 2.3 $\mathrm{MeV}$. The CDF Collaboration update the mass of $X(3872)$ as
\bqa
M(X(3872))=3871.61\pm 0.16\pm 0.19 \mathrm{MeV},
\eqa
which is just below the $D^0\bar{D}^{*0}$ threshold $m(D^0\bar{D}^{*0})=3871.81\pm 0.36 \mathrm{MeV}$.
Since this state is unexpected in the quenched potential model, its
origin was widely discussed based on a molecule candidate of
$D^0\bar{D}^{*0}$, a tetro-quark state, a charmonium state, or a
charmonium state mixed with the $D^0\bar{D}^{*0}$
component~(See Ref.\cite{Brambilla:2010cs} and the references therein).
Our calculation here supports the idea to regard $X(3872)$ as a
charmonium state mixed with the $D^0\bar{D}^{*0}$ component, which is
described in the language of hadron loops here. The bare mass of
$\chi_{c1}(2^3P_1)$ is at 3950 $\mathrm{MeV}$ in the GI's model
prediction, while its coupling to the $D^0\bar{D}^{*0}$ and
$D^+\bar{D}^{*-}$ thresholds reduces its pole mass to 3884 $\mathrm{MeV}$ based on the parameter set we choose. The pole mass is just about 12 $\mathrm{MeV}$ higher than the observed mass of the $X(3872)$, and the related pole width is about 3 $\mathrm{MeV}$, which is comparable with the experimental data.

\begin{figure}[t]%
\begin{center}%
\includegraphics[height=50mm]{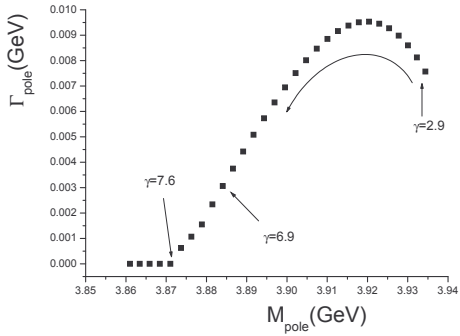}
\caption{\label{trj3872} The trajectory of the pole parameters of the $\chi_{c1}(2P)$ state when the $\gamma$ parameter increases.}
\end{center}%
\end{figure}

Moreover, it is worth pointing out that the $\gamma$ parameter of the
QPC model usually has an uncertainty of about
$30\%$~\cite{PhysRevD.72.094004}, which is determined by fitting to
the decay experiment data.  The
pole trajectory of $\chi_{c1}(2P)$ state is shown in
Fig.\ref{trj3872}, with the $\gamma$ parameter ranging from $2.9$ to
$8.9$.  When this coupling is very
weak, the pole of the $\chi_{c1}(2P)$ state is close to the bare
state on the real energy axis, whose mass is $3950 \mathrm{MeV}$ predicted in the GI's model.
 Along with the strengthening of this coupling, the pole mass is pulled down with its width always below
$10 \mathrm{MeV}$. When the coupling become stronger at around
$\gamma=7.6$, which is just $10\%$ away from the central value, the
pole will reach the $D^0\bar{D}^{*0}$ threshold and then becomes a
virtual bound state.  Since the pole is shifted from the bare mass of
$\chi_{c1}(2P)$, a charmonium origin of $X(3872)$ is easily suggested,
or the $X(3872)$ could be regarded as a charmonium state dressed by
the $D\bar{D}^*$ cloud. Such a mixed state is as compact as a
conventional charmonium state, and this assignment could resolve the
problems encountered by the molecule description in explaining the
radiative transition $Br(X\rightarrow \psi'\gamma)/ Br(X\rightarrow
\psi\gamma)$ ratio~\cite{Aubert:2008ae}.
This picture has some similarities with the other
coupled channel analyses in  Ref.\cite{Danilkin:2010cc} and
Ref.~\cite{Coito:2010if}, but we wish to point out that our
calculation  considers not only the $2P$ states but the
charmonium-like spectrum systematically. Additionally, the authors of
\cite{Zhang:2009bv} used a coupled channel Flatt$\acute{\mathrm{e}}$ formula to fit the experimental data, and
suggested that there may need to be two near-threshold poles to
account for the data, one from the $D^0\bar{D}^{*0}$ component and the
other from the charmonium state $\chi_{c1}(2P)$.

In the recoiling spectrum of $J/\psi$ in the $e^+e^-$ annihilation process $e^+e^-\rightarrow J/\psi+D\bar{D}^*$,  the Belle group~\cite{Abe:2007jna,Abe:2007sya} found the evidence of the $X(3940)$, whose mass and width are determined as
\bqa
M(X(3940))&=&3942^{+7}_{-6}\pm 6 \mathrm{MeV},\nonumber\\
\Gamma(X(3940))&=&37^{+26}_{-18}\pm 8 \mathrm{MeV}.
\eqa
Meanwhile, they also found the $X(4160)$ in the $D^*\bar{D}^*$ mode in the process $e^+e^-\rightarrow J/\psi+D^*\bar{D}^*$. The mass and width of $X(4160)$ are given by
\bqa
M(X(4160))&=&4156^{+25}_{-20}\pm 15 \mathrm{MeV},\nonumber\\
\Gamma(X(4160))&=&139^{+111}_{-61}\pm 21 \mathrm{MeV}.
\eqa
Besides, there is a broad enhancement around $3880 \mathrm{MeV}$ in
the $D\bar{D}$ spectrum in $e^+e^-\rightarrow J/\psi+D\bar{D}$. Even
though in Ref.~\cite{Abe:2007jna} the authors regard this structure as too wide to present a resonance shape
sufficiently, the similar evidence was also reported by the BaBar
group.~\cite{Aubert:2006mi} The charge parities of the two $X$-states
are suggested to be even since the charge odd state associated with
$J/\psi$ needs to be produced via two photon fragmentation, which is
expected to be highly suppressed~\cite{Chao:2007it}. Here, one can
find that the pole parameters of $h_c(2^1P_1)$ and $\chi_0(3^3P_0)$
naturally fit in with $X(3940)$ and $X(4160)$ respectively.
Furthermore, the broad structure in the $D\bar{D}$
spectrum~\cite{Abe:2007jna,Aubert:2006mi} could be the $\chi_{c0}(2P)$
state, since its mass in this calculation lies at the nearby location
with also a fairly large width $133 \mathrm{MeV}$. There are several
difficulties in assigning the $X(3915)$ as $\chi_{c0}(2P)$ as
discussed in Ref.~\cite{Guo:2012tv}. The authors of
Ref.~\cite{Guo:2012tv} also made a fit to the data of
$\gamma\gamma\rightarrow D\bar{D}$ of Belle~\cite{Uehara:2005qd} and
BaBar~\cite{Aubert:2010ab}, which also indicates a broad structure
with a mass $M=3837\pm 12 \mathrm{MeV}$ and a width $\Gamma=221\pm 19 \mathrm{MeV}$.  Further experiments are required for clarifying this issue.

\section{Summary}

In this calculation, we try to incorporate the hadron loop effect, due
to the light quark pair creation from the vacuum, to investigate the
charmonium-like spectrum and the decays of its members in one unified
picture. We calculate the pole masses and widths of those
charmonium-like states above the $D\bar{D}$ threshold and give
possible assignments for the newly-observed $\psi$-like or $X$ states.
The hadron loop effect generally shifts the pole mass of a state
down from its mass predicted in the potential model. These shifts are
helpful in our understanding of most observed states.
Typically, the pole mass of $\chi_{c1}(2P)$ could be lowered
significantly and reach the region of  $X(3872)$, which implies that
the proximity of  $X(3872)$ to the $D^0\bar{D}^{*0}$ might be an
accident due to its coupling to the nearby $D\bar{D}^{*}$ thresholds.
This state could have a charmonium origin with a few percent
$D\bar{D}^{*}$ components. The pole mass of $\psi(4S)$ state has also
a shift of about 100 $\mathrm{MeV}$, whose value is more compatible
with the $X(4360)$ but not with the $\psi(4415)$. We also point out
that the $\chi_{c0}(2P)$ state
could probably be a broad state at
about  3880 $\mathrm{MeV}$, but not the narrow $X(3915)$. It  requires
further theoretical and experimental explorations to clarify the
nature of $X(3915)$.

\begin{acknowledgments}
Z.Z. would like to thank the Project Sponsored by the Scientific Research Foundation for the Returned Overseas Chinese Scholars, State Education Ministry. Z.X. is supported by the National Natural Science Foundation of China under grant No.11105138 and
11235010 and the Fundamental Research Funds for the
Central Universities under grant No.WK2030040020.

\end{acknowledgments}

\bibliographystyle{apsrev4-1}
\bibliography{charmlike}

\begin{thebibliography}{51}%
\makeatletter
\providecommand \@ifxundefined [1]{%
 \@ifx{#1\undefined}
}%
\providecommand \@ifnum [1]{%
 \ifnum #1\expandafter \@firstoftwo
 \else \expandafter \@secondoftwo
 \fi
}%
\providecommand \@ifx [1]{%
 \ifx #1\expandafter \@firstoftwo
 \else \expandafter \@secondoftwo
 \fi
}%
\providecommand \natexlab [1]{#1}%
\providecommand \enquote  [1]{``#1''}%
\providecommand \bibnamefont  [1]{#1}%
\providecommand \bibfnamefont [1]{#1}%
\providecommand \citenamefont [1]{#1}%
\providecommand \href@noop [0]{\@secondoftwo}%
\providecommand \href [0]{\begingroup \@sanitize@url \@href}%
\providecommand \@href[1]{\@@startlink{#1}\@@href}%
\providecommand \@@href[1]{\endgroup#1\@@endlink}%
\providecommand \@sanitize@url [0]{\catcode `\\12\catcode `\$12\catcode
  `\&12\catcode `\#12\catcode `\^12\catcode `\_12\catcode `\%12\relax}%
\providecommand \@@startlink[1]{}%
\providecommand \@@endlink[0]{}%
\providecommand \url  [0]{\begingroup\@sanitize@url \@url }%
\providecommand \@url [1]{\endgroup\@href {#1}{\urlprefix }}%
\providecommand \urlprefix  [0]{URL }%
\providecommand \Eprint [0]{\href }%
\providecommand \doibase [0]{http://dx.doi.org/}%
\providecommand \selectlanguage [0]{\@gobble}%
\providecommand \bibinfo  [0]{\@secondoftwo}%
\providecommand \bibfield  [0]{\@secondoftwo}%
\providecommand \translation [1]{[#1]}%
\providecommand \BibitemOpen [0]{}%
\providecommand \bibitemStop [0]{}%
\providecommand \bibitemNoStop [0]{.\EOS\space}%
\providecommand \EOS [0]{\spacefactor3000\relax}%
\providecommand \BibitemShut  [1]{\csname bibitem#1\endcsname}%
\let\auto@bib@innerbib\@empty
\bibitem [{\citenamefont {Choi}\ \emph {et~al.}(2003)\citenamefont {Choi} \emph
  {et~al.}}]{Choi:2003ue}%
  \BibitemOpen
  \bibfield  {author} {\bibinfo {author} {\bibfnamefont {S.}~\bibnamefont
  {Choi}} \emph {et~al.} (\bibinfo {collaboration} {Belle Collaboration}),\
  }\href {\doibase 10.1103/PhysRevLett.91.262001} {\bibfield  {journal}
  {\bibinfo  {journal} {Phys.Rev.Lett.}\ }\textbf {\bibinfo {volume} {91}},\
  \bibinfo {pages} {262001} (\bibinfo {year} {2003})},\ \Eprint
  {http://arxiv.org/abs/hep-ex/0309032} {arXiv:hep-ex/0309032 [hep-ex]}
  \BibitemShut {NoStop}%
\bibitem [{\citenamefont {Beringer}\ \emph {et~al.}(2012)\citenamefont
  {Beringer} \emph {et~al.}}]{Beringer:1900zz}%
  \BibitemOpen
  \bibfield  {author} {\bibinfo {author} {\bibfnamefont {J.}~\bibnamefont
  {Beringer}} \emph {et~al.} (\bibinfo {collaboration} {Particle Data Group}),\
  }\href {\doibase 10.1103/PhysRevD.86.010001} {\bibfield  {journal} {\bibinfo
  {journal} {Phys.Rev.}\ }\textbf {\bibinfo {volume} {D86}},\ \bibinfo {pages}
  {010001} (\bibinfo {year} {2012})}\BibitemShut {NoStop}%
\bibitem [{\citenamefont {Godfrey}\ and\ \citenamefont
  {Isgur}(1985)}]{Godfrey:1985xj}%
  \BibitemOpen
  \bibfield  {author} {\bibinfo {author} {\bibfnamefont {S.}~\bibnamefont
  {Godfrey}}\ and\ \bibinfo {author} {\bibfnamefont {N.}~\bibnamefont
  {Isgur}},\ }\href {\doibase 10.1103/PhysRevD.32.189} {\bibfield  {journal}
  {\bibinfo  {journal} {Phys. Rev.}\ }\textbf {\bibinfo {volume} {D32}},\
  \bibinfo {pages} {189} (\bibinfo {year} {1985})}\BibitemShut {NoStop}%
\bibitem [{\citenamefont {Barnes}\ and\ \citenamefont
  {Godfrey}(2004)}]{Barnes:2003vb}%
  \BibitemOpen
  \bibfield  {author} {\bibinfo {author} {\bibfnamefont {T.}~\bibnamefont
  {Barnes}}\ and\ \bibinfo {author} {\bibfnamefont {S.}~\bibnamefont
  {Godfrey}},\ }\href {\doibase 10.1103/PhysRevD.69.054008} {\bibfield
  {journal} {\bibinfo  {journal} {Phys.Rev.}\ }\textbf {\bibinfo {volume}
  {D69}},\ \bibinfo {pages} {054008} (\bibinfo {year} {2004})},\ \Eprint
  {http://arxiv.org/abs/hep-ph/0311162} {arXiv:hep-ph/0311162 [hep-ph]}
  \BibitemShut {NoStop}%
\bibitem [{\citenamefont {Tornqvist}(2004)}]{Tornqvist:2004qy}%
  \BibitemOpen
  \bibfield  {author} {\bibinfo {author} {\bibfnamefont {N.~A.}\ \bibnamefont
  {Tornqvist}},\ }\href {\doibase 10.1016/j.physletb.2004.03.077} {\bibfield
  {journal} {\bibinfo  {journal} {Phys.Lett.}\ }\textbf {\bibinfo {volume}
  {B590}},\ \bibinfo {pages} {209} (\bibinfo {year} {2004})},\ \Eprint
  {http://arxiv.org/abs/hep-ph/0402237} {arXiv:hep-ph/0402237 [hep-ph]}
  \BibitemShut {NoStop}%
\bibitem [{\citenamefont {Close}\ and\ \citenamefont
  {Page}(2004)}]{Close:2003sg}%
  \BibitemOpen
  \bibfield  {author} {\bibinfo {author} {\bibfnamefont {F.~E.}\ \bibnamefont
  {Close}}\ and\ \bibinfo {author} {\bibfnamefont {P.~R.}\ \bibnamefont
  {Page}},\ }\href {\doibase 10.1016/j.physletb.2003.10.032} {\bibfield
  {journal} {\bibinfo  {journal} {Phys.Lett.}\ }\textbf {\bibinfo {volume}
  {B578}},\ \bibinfo {pages} {119} (\bibinfo {year} {2004})},\ \Eprint
  {http://arxiv.org/abs/hep-ph/0309253} {arXiv:hep-ph/0309253 [hep-ph]}
  \BibitemShut {NoStop}%
\bibitem [{\citenamefont {Voloshin}(2004)}]{Voloshin:2003nt}%
  \BibitemOpen
  \bibfield  {author} {\bibinfo {author} {\bibfnamefont {M.}~\bibnamefont
  {Voloshin}},\ }\href {\doibase 10.1016/j.physletb.2003.11.014} {\bibfield
  {journal} {\bibinfo  {journal} {Phys.Lett.}\ }\textbf {\bibinfo {volume}
  {B579}},\ \bibinfo {pages} {316} (\bibinfo {year} {2004})},\ \Eprint
  {http://arxiv.org/abs/hep-ph/0309307} {arXiv:hep-ph/0309307 [hep-ph]}
  \BibitemShut {NoStop}%
\bibitem [{\citenamefont {Wong}(2004)}]{Wong:2003xk}%
  \BibitemOpen
  \bibfield  {author} {\bibinfo {author} {\bibfnamefont {C.-Y.}\ \bibnamefont
  {Wong}},\ }\href {\doibase 10.1103/PhysRevC.69.055202} {\bibfield  {journal}
  {\bibinfo  {journal} {Phys.Rev.}\ }\textbf {\bibinfo {volume} {C69}},\
  \bibinfo {pages} {055202} (\bibinfo {year} {2004})},\ \Eprint
  {http://arxiv.org/abs/hep-ph/0311088} {arXiv:hep-ph/0311088 [hep-ph]}
  \BibitemShut {NoStop}%
\bibitem [{\citenamefont {Braaten}\ and\ \citenamefont
  {Kusunoki}(2004)}]{Braaten:2003he}%
  \BibitemOpen
  \bibfield  {author} {\bibinfo {author} {\bibfnamefont {E.}~\bibnamefont
  {Braaten}}\ and\ \bibinfo {author} {\bibfnamefont {M.}~\bibnamefont
  {Kusunoki}},\ }\href {\doibase 10.1103/PhysRevD.69.074005} {\bibfield
  {journal} {\bibinfo  {journal} {Phys.Rev.}\ }\textbf {\bibinfo {volume}
  {D69}},\ \bibinfo {pages} {074005} (\bibinfo {year} {2004})},\ \Eprint
  {http://arxiv.org/abs/hep-ph/0311147} {arXiv:hep-ph/0311147 [hep-ph]}
  \BibitemShut {NoStop}%
\bibitem [{\citenamefont {Aubert}\ \emph {et~al.}(2009)\citenamefont {Aubert}
  \emph {et~al.}}]{Aubert:2008ae}%
  \BibitemOpen
  \bibfield  {author} {\bibinfo {author} {\bibfnamefont {B.}~\bibnamefont
  {Aubert}} \emph {et~al.} (\bibinfo {collaboration} {BaBar Collaboration}),\
  }\href {\doibase 10.1103/PhysRevLett.102.132001} {\bibfield  {journal}
  {\bibinfo  {journal} {Phys.Rev.Lett.}\ }\textbf {\bibinfo {volume} {102}},\
  \bibinfo {pages} {132001} (\bibinfo {year} {2009})},\ \Eprint
  {http://arxiv.org/abs/0809.0042} {arXiv:0809.0042 [hep-ex]} \BibitemShut
  {NoStop}%
\bibitem [{\citenamefont {Swanson}(2004{\natexlab{a}})}]{Swanson:2003tb}%
  \BibitemOpen
  \bibfield  {author} {\bibinfo {author} {\bibfnamefont {E.~S.}\ \bibnamefont
  {Swanson}},\ }\href {\doibase 10.1016/j.physletb.2004.03.033} {\bibfield
  {journal} {\bibinfo  {journal} {Phys.Lett.}\ }\textbf {\bibinfo {volume}
  {B588}},\ \bibinfo {pages} {189} (\bibinfo {year} {2004}{\natexlab{a}})},\
  \Eprint {http://arxiv.org/abs/hep-ph/0311229} {arXiv:hep-ph/0311229 [hep-ph]}
  \BibitemShut {NoStop}%
\bibitem [{\citenamefont {Swanson}(2004{\natexlab{b}})}]{Swanson:2004pp}%
  \BibitemOpen
  \bibfield  {author} {\bibinfo {author} {\bibfnamefont {E.~S.}\ \bibnamefont
  {Swanson}},\ }\href {\doibase 10.1016/j.physletb.2004.07.059} {\bibfield
  {journal} {\bibinfo  {journal} {Phys.Lett.}\ }\textbf {\bibinfo {volume}
  {B598}},\ \bibinfo {pages} {197} (\bibinfo {year} {2004}{\natexlab{b}})},\
  \Eprint {http://arxiv.org/abs/hep-ph/0406080} {arXiv:hep-ph/0406080 [hep-ph]}
  \BibitemShut {NoStop}%
\bibitem [{\citenamefont {Eichten}\ \emph {et~al.}(1978)\citenamefont
  {Eichten}, \citenamefont {Gottfried}, \citenamefont {Kinoshita},
  \citenamefont {Lane},\ and\ \citenamefont {Yan}}]{Eichten:1978tg}%
  \BibitemOpen
  \bibfield  {author} {\bibinfo {author} {\bibfnamefont {E.}~\bibnamefont
  {Eichten}}, \bibinfo {author} {\bibfnamefont {K.}~\bibnamefont {Gottfried}},
  \bibinfo {author} {\bibfnamefont {T.}~\bibnamefont {Kinoshita}}, \bibinfo
  {author} {\bibfnamefont {K.}~\bibnamefont {Lane}}, \ and\ \bibinfo {author}
  {\bibfnamefont {T.-M.}\ \bibnamefont {Yan}},\ }\href {\doibase
  10.1103/PhysRevD.17.3090, 10.1103/PhysRevD.21.313} {\bibfield  {journal}
  {\bibinfo  {journal} {Phys.Rev.}\ }\textbf {\bibinfo {volume} {D17}},\
  \bibinfo {pages} {3090} (\bibinfo {year} {1978})}\BibitemShut {NoStop}%
\bibitem [{\citenamefont {van Beveren}\ \emph {et~al.}(1983)\citenamefont {van
  Beveren}, \citenamefont {Dullemond},\ and\ \citenamefont
  {Rijken}}]{vanBeveren:1983td}%
  \BibitemOpen
  \bibfield  {author} {\bibinfo {author} {\bibfnamefont {E.}~\bibnamefont {van
  Beveren}}, \bibinfo {author} {\bibfnamefont {C.}~\bibnamefont {Dullemond}}, \
  and\ \bibinfo {author} {\bibfnamefont {T.}~\bibnamefont {Rijken}},\ }\href
  {\doibase 10.1007/BF01572256} {\bibfield  {journal} {\bibinfo  {journal}
  {Z.Phys.}\ }\textbf {\bibinfo {volume} {C19}},\ \bibinfo {pages} {275}
  (\bibinfo {year} {1983})}\BibitemShut {NoStop}%
\bibitem [{\citenamefont {Heikkila}\ \emph {et~al.}(1984)\citenamefont
  {Heikkila}, \citenamefont {Tornqvist},\ and\ \citenamefont
  {Ono}}]{Heikkila:1983wd}%
  \BibitemOpen
  \bibfield  {author} {\bibinfo {author} {\bibfnamefont {K.}~\bibnamefont
  {Heikkila}}, \bibinfo {author} {\bibfnamefont {N.~A.}\ \bibnamefont
  {Tornqvist}}, \ and\ \bibinfo {author} {\bibfnamefont {S.}~\bibnamefont
  {Ono}},\ }\href {\doibase 10.1103/PhysRevD.29.110} {\bibfield  {journal}
  {\bibinfo  {journal} {Phys. Rev.}\ }\textbf {\bibinfo {volume} {D29}},\
  \bibinfo {pages} {110} (\bibinfo {year} {1984})}\BibitemShut {NoStop}%
\bibitem [{\citenamefont {Pennington}\ and\ \citenamefont
  {Wilson}(2007)}]{Pennington:2007xr}%
  \BibitemOpen
  \bibfield  {author} {\bibinfo {author} {\bibfnamefont {M.~R.}\ \bibnamefont
  {Pennington}}\ and\ \bibinfo {author} {\bibfnamefont {D.~J.}\ \bibnamefont
  {Wilson}},\ }\href {\doibase 10.1103/PhysRevD.76.077502} {\bibfield
  {journal} {\bibinfo  {journal} {Phys. Rev.}\ }\textbf {\bibinfo {volume}
  {D76}},\ \bibinfo {pages} {077502} (\bibinfo {year} {2007})},\ \Eprint
  {http://arxiv.org/abs/0704.3384} {arXiv:0704.3384 [hep-ph]} \BibitemShut
  {NoStop}%
\bibitem [{\citenamefont {Barnes}\ \emph {et~al.}(2005)\citenamefont {Barnes},
  \citenamefont {Godfrey},\ and\ \citenamefont {Swanson}}]{Barnes:2005pb}%
  \BibitemOpen
  \bibfield  {author} {\bibinfo {author} {\bibfnamefont {T.}~\bibnamefont
  {Barnes}}, \bibinfo {author} {\bibfnamefont {S.}~\bibnamefont {Godfrey}}, \
  and\ \bibinfo {author} {\bibfnamefont {E.}~\bibnamefont {Swanson}},\ }\href
  {\doibase 10.1103/PhysRevD.72.054026} {\bibfield  {journal} {\bibinfo
  {journal} {Phys.Rev.}\ }\textbf {\bibinfo {volume} {D72}},\ \bibinfo {pages}
  {054026} (\bibinfo {year} {2005})},\ \Eprint
  {http://arxiv.org/abs/hep-ph/0505002} {arXiv:hep-ph/0505002 [hep-ph]}
  \BibitemShut {NoStop}%
\bibitem [{\citenamefont {Chao}\ and\ \citenamefont {Liu}(1989)}]{Chao:1989ek}%
  \BibitemOpen
  \bibfield  {author} {\bibinfo {author} {\bibfnamefont {K.-T.}\ \bibnamefont
  {Chao}}\ and\ \bibinfo {author} {\bibfnamefont {J.-H.}\ \bibnamefont {Liu}},\
  }\href@noop {} {\bibfield  {journal} {\bibinfo  {journal} {Proceedings of the
  Workshop on Weak Interactions and CP Violation, Beijing, August 22-26, 1989,
  edited by T. Huang and D.D. Wu, World Scientific (Singapore, 1990)
  p.109-p.117}\ } (\bibinfo {year} {1989})}\BibitemShut {NoStop}%
\bibitem [{\citenamefont {Li}\ and\ \citenamefont {Chao}(2009)}]{Li:2009zu}%
  \BibitemOpen
  \bibfield  {author} {\bibinfo {author} {\bibfnamefont {B.-Q.}\ \bibnamefont
  {Li}}\ and\ \bibinfo {author} {\bibfnamefont {K.-T.}\ \bibnamefont {Chao}},\
  }\href {\doibase 10.1103/PhysRevD.79.094004} {\bibfield  {journal} {\bibinfo
  {journal} {Phys.Rev.}\ }\textbf {\bibinfo {volume} {D79}},\ \bibinfo {pages}
  {094004} (\bibinfo {year} {2009})},\ \Eprint {http://arxiv.org/abs/0903.5506}
  {arXiv:0903.5506 [hep-ph]} \BibitemShut {NoStop}%
\bibitem [{\citenamefont {Pennington}(2010)}]{Pennington:2010dc}%
  \BibitemOpen
  \bibfield  {author} {\bibinfo {author} {\bibfnamefont {M.~R.}\ \bibnamefont
  {Pennington}},\ }\href {\doibase 10.1063/1.3483333} {\bibfield  {journal}
  {\bibinfo  {journal} {AIP Conf. Proc.}\ }\textbf {\bibinfo {volume} {1257}},\
  \bibinfo {pages} {27} (\bibinfo {year} {2010})},\ \Eprint
  {http://arxiv.org/abs/1003.2549} {arXiv:1003.2549 [hep-ph]} \BibitemShut
  {NoStop}%
\bibitem [{\citenamefont {Geiger}\ and\ \citenamefont
  {Isgur}(1993)}]{Geiger:1992va}%
  \BibitemOpen
  \bibfield  {author} {\bibinfo {author} {\bibfnamefont {P.}~\bibnamefont
  {Geiger}}\ and\ \bibinfo {author} {\bibfnamefont {N.}~\bibnamefont {Isgur}},\
  }\href {\doibase 10.1103/PhysRevD.47.5050} {\bibfield  {journal} {\bibinfo
  {journal} {Phys. Rev.}\ }\textbf {\bibinfo {volume} {D47}},\ \bibinfo {pages}
  {5050} (\bibinfo {year} {1993})}\BibitemShut {NoStop}%
\bibitem [{\citenamefont {Micu}(1969)}]{Micu:1968mk}%
  \BibitemOpen
  \bibfield  {author} {\bibinfo {author} {\bibfnamefont {L.}~\bibnamefont
  {Micu}},\ }\href {\doibase 10.1016/0550-3213(69)90039-X} {\bibfield
  {journal} {\bibinfo  {journal} {Nucl. Phys.}\ }\textbf {\bibinfo {volume}
  {B10}},\ \bibinfo {pages} {521} (\bibinfo {year} {1969})}\BibitemShut
  {NoStop}%
\bibitem [{\citenamefont {Colglazier}\ and\ \citenamefont
  {Rosner}(1971)}]{Colglazier}%
  \BibitemOpen
  \bibfield  {author} {\bibinfo {author} {\bibfnamefont {E.~W.}\ \bibnamefont
  {Colglazier}}\ and\ \bibinfo {author} {\bibfnamefont {J.~L.}\ \bibnamefont
  {Rosner}},\ }\href {\doibase 10.1016/0550-3213(71)90100-3} {\bibfield
  {journal} {\bibinfo  {journal} {Nucl. Phys.}\ }\textbf {\bibinfo {volume}
  {B27}},\ \bibinfo {pages} {349} (\bibinfo {year} {1971})}\BibitemShut
  {NoStop}%
\bibitem [{\citenamefont {Le~Yaouanc}\ \emph {et~al.}(1973)\citenamefont
  {Le~Yaouanc}, \citenamefont {Oliver}, \citenamefont {Pene},\ and\
  \citenamefont {Raynal}}]{LeYaouanc:1972ae}%
  \BibitemOpen
  \bibfield  {author} {\bibinfo {author} {\bibfnamefont {A.}~\bibnamefont
  {Le~Yaouanc}}, \bibinfo {author} {\bibfnamefont {L.}~\bibnamefont {Oliver}},
  \bibinfo {author} {\bibfnamefont {O.}~\bibnamefont {Pene}}, \ and\ \bibinfo
  {author} {\bibfnamefont {J.~C.}\ \bibnamefont {Raynal}},\ }\href {\doibase
  10.1103/PhysRevD.8.2223} {\bibfield  {journal} {\bibinfo  {journal} {Phys.
  Rev.}\ }\textbf {\bibinfo {volume} {D8}},\ \bibinfo {pages} {2223} (\bibinfo
  {year} {1973})}\BibitemShut {NoStop}%
\bibitem [{\citenamefont {Luo}\ \emph {et~al.}(2009)\citenamefont {Luo},
  \citenamefont {Chen},\ and\ \citenamefont {Liu}}]{Luo:2009wu}%
  \BibitemOpen
  \bibfield  {author} {\bibinfo {author} {\bibfnamefont {Z.-G.}\ \bibnamefont
  {Luo}}, \bibinfo {author} {\bibfnamefont {X.-L.}\ \bibnamefont {Chen}}, \
  and\ \bibinfo {author} {\bibfnamefont {X.}~\bibnamefont {Liu}},\ }\href
  {\doibase 10.1103/PhysRevD.79.074020} {\bibfield  {journal} {\bibinfo
  {journal} {Phys.Rev.}\ }\textbf {\bibinfo {volume} {D79}},\ \bibinfo {pages}
  {074020} (\bibinfo {year} {2009})},\ \Eprint {http://arxiv.org/abs/0901.0505}
  {arXiv:0901.0505 [hep-ph]} \BibitemShut {NoStop}%
\bibitem [{\citenamefont {Eden}\ and\ \citenamefont
  {Taylor}(1964)}]{Eden:1964zz}%
  \BibitemOpen
  \bibfield  {author} {\bibinfo {author} {\bibfnamefont {R.~J.}\ \bibnamefont
  {Eden}}\ and\ \bibinfo {author} {\bibfnamefont {J.~R.}\ \bibnamefont
  {Taylor}},\ }\href {\doibase 10.1103/PhysRev.133.B1575} {\bibfield  {journal}
  {\bibinfo  {journal} {Phys. Rev.}\ }\textbf {\bibinfo {volume} {133}},\
  \bibinfo {pages} {B1575} (\bibinfo {year} {1964})}\BibitemShut {NoStop}%
\bibitem [{\citenamefont {Godfrey}\ and\ \citenamefont
  {Kokoski}(1991)}]{Godfrey:1986wj}%
  \BibitemOpen
  \bibfield  {author} {\bibinfo {author} {\bibfnamefont {S.}~\bibnamefont
  {Godfrey}}\ and\ \bibinfo {author} {\bibfnamefont {R.}~\bibnamefont
  {Kokoski}},\ }\href {\doibase 10.1103/PhysRevD.43.1679} {\bibfield  {journal}
  {\bibinfo  {journal} {Phys. Rev.}\ }\textbf {\bibinfo {volume} {D43}},\
  \bibinfo {pages} {1679} (\bibinfo {year} {1991})}\BibitemShut {NoStop}%
\bibitem [{\citenamefont {Kokoski}\ and\ \citenamefont
  {Isgur}(1987)}]{Kokoski:1985is}%
  \BibitemOpen
  \bibfield  {author} {\bibinfo {author} {\bibfnamefont {R.}~\bibnamefont
  {Kokoski}}\ and\ \bibinfo {author} {\bibfnamefont {N.}~\bibnamefont
  {Isgur}},\ }\href {\doibase 10.1103/PhysRevD.35.907} {\bibfield  {journal}
  {\bibinfo  {journal} {Phys. Rev.}\ }\textbf {\bibinfo {volume} {D35}},\
  \bibinfo {pages} {907} (\bibinfo {year} {1987})}\BibitemShut {NoStop}%
\bibitem [{\citenamefont {Close}\ and\ \citenamefont
  {Swanson}(2005{\natexlab{a}})}]{Close:2005se}%
  \BibitemOpen
  \bibfield  {author} {\bibinfo {author} {\bibfnamefont {F.}~\bibnamefont
  {Close}}\ and\ \bibinfo {author} {\bibfnamefont {E.}~\bibnamefont
  {Swanson}},\ }\href {\doibase 10.1103/PhysRevD.72.094004} {\bibfield
  {journal} {\bibinfo  {journal} {Phys.Rev.}\ }\textbf {\bibinfo {volume}
  {D72}},\ \bibinfo {pages} {094004} (\bibinfo {year} {2005}{\natexlab{a}})},\
  \Eprint {http://arxiv.org/abs/hep-ph/0505206} {arXiv:hep-ph/0505206 [hep-ph]}
  \BibitemShut {NoStop}%
\bibitem [{\citenamefont {Le~Yaouanc}\ \emph {et~al.}()\citenamefont
  {Le~Yaouanc}, \citenamefont {Oliver}, \citenamefont {Pene},\ and\
  \citenamefont {Raynal}}]{yaouancbook}%
  \BibitemOpen
  \bibfield  {author} {\bibinfo {author} {\bibfnamefont {A.}~\bibnamefont
  {Le~Yaouanc}}, \bibinfo {author} {\bibfnamefont {L.}~\bibnamefont {Oliver}},
  \bibinfo {author} {\bibfnamefont {O.}~\bibnamefont {Pene}}, \ and\ \bibinfo
  {author} {\bibfnamefont {J.~C.}\ \bibnamefont {Raynal}},\ }\href@noop {} {\
  }\bibinfo {note} {NEW YORK, USA: GORDON AND BREACH (1988) 311p}\BibitemShut
  {NoStop}%
\bibitem [{\citenamefont {Rosner}(2005)}]{Rosner:2004wy}%
  \BibitemOpen
  \bibfield  {author} {\bibinfo {author} {\bibfnamefont {J.~L.}\ \bibnamefont
  {Rosner}},\ }\href {\doibase 10.1016/j.aop.2005.02.004} {\bibfield  {journal}
  {\bibinfo  {journal} {Annals Phys.}\ }\textbf {\bibinfo {volume} {319}},\
  \bibinfo {pages} {1} (\bibinfo {year} {2005})},\ \Eprint
  {http://arxiv.org/abs/hep-ph/0411003} {arXiv:hep-ph/0411003 [hep-ph]}
  \BibitemShut {NoStop}%
\bibitem [{\citenamefont {Aubert}\ \emph {et~al.}(2005)\citenamefont {Aubert}
  \emph {et~al.}}]{Aubert:2005rm}%
  \BibitemOpen
  \bibfield  {author} {\bibinfo {author} {\bibfnamefont {B.}~\bibnamefont
  {Aubert}} \emph {et~al.} (\bibinfo {collaboration} {BaBar Collaboration}),\
  }\href {\doibase 10.1103/PhysRevLett.95.142001} {\bibfield  {journal}
  {\bibinfo  {journal} {Phys.Rev.Lett.}\ }\textbf {\bibinfo {volume} {95}},\
  \bibinfo {pages} {142001} (\bibinfo {year} {2005})},\ \Eprint
  {http://arxiv.org/abs/hep-ex/0506081} {arXiv:hep-ex/0506081 [hep-ex]}
  \BibitemShut {NoStop}%
\bibitem [{\citenamefont {Zhu}(2005)}]{Zhu:2005hp}%
  \BibitemOpen
  \bibfield  {author} {\bibinfo {author} {\bibfnamefont {S.-L.}\ \bibnamefont
  {Zhu}},\ }\href {\doibase 10.1016/j.physletb.2005.08.068} {\bibfield
  {journal} {\bibinfo  {journal} {Phys.Lett.}\ }\textbf {\bibinfo {volume}
  {B625}},\ \bibinfo {pages} {212} (\bibinfo {year} {2005})},\ \Eprint
  {http://arxiv.org/abs/hep-ph/0507025} {arXiv:hep-ph/0507025 [hep-ph]}
  \BibitemShut {NoStop}%
\bibitem [{\citenamefont {Maiani}\ \emph {et~al.}(2005)\citenamefont {Maiani},
  \citenamefont {Riquer}, \citenamefont {Piccinini},\ and\ \citenamefont
  {Polosa}}]{Maiani:2005pe}%
  \BibitemOpen
  \bibfield  {author} {\bibinfo {author} {\bibfnamefont {L.}~\bibnamefont
  {Maiani}}, \bibinfo {author} {\bibfnamefont {V.}~\bibnamefont {Riquer}},
  \bibinfo {author} {\bibfnamefont {F.}~\bibnamefont {Piccinini}}, \ and\
  \bibinfo {author} {\bibfnamefont {A.}~\bibnamefont {Polosa}},\ }\href
  {\doibase 10.1103/PhysRevD.72.031502} {\bibfield  {journal} {\bibinfo
  {journal} {Phys.Rev.}\ }\textbf {\bibinfo {volume} {D72}},\ \bibinfo {pages}
  {031502} (\bibinfo {year} {2005})},\ \Eprint
  {http://arxiv.org/abs/hep-ph/0507062} {arXiv:hep-ph/0507062 [hep-ph]}
  \BibitemShut {NoStop}%
\bibitem [{\citenamefont {Ebert}\ \emph {et~al.}(2008)\citenamefont {Ebert},
  \citenamefont {Faustov},\ and\ \citenamefont {Galkin}}]{Ebert:2008kb}%
  \BibitemOpen
  \bibfield  {author} {\bibinfo {author} {\bibfnamefont {D.}~\bibnamefont
  {Ebert}}, \bibinfo {author} {\bibfnamefont {R.}~\bibnamefont {Faustov}}, \
  and\ \bibinfo {author} {\bibfnamefont {V.}~\bibnamefont {Galkin}},\ }\href
  {\doibase 10.1140/epjc/s10052-008-0754-8} {\bibfield  {journal} {\bibinfo
  {journal} {Eur.Phys.J.}\ }\textbf {\bibinfo {volume} {C58}},\ \bibinfo
  {pages} {399} (\bibinfo {year} {2008})},\ \Eprint
  {http://arxiv.org/abs/0808.3912} {arXiv:0808.3912 [hep-ph]} \BibitemShut
  {NoStop}%
\bibitem [{\citenamefont {Albuquerque}\ and\ \citenamefont
  {Nielsen}(2009)}]{Albuquerque:2008up}%
  \BibitemOpen
  \bibfield  {author} {\bibinfo {author} {\bibfnamefont {R.}~\bibnamefont
  {Albuquerque}}\ and\ \bibinfo {author} {\bibfnamefont {M.}~\bibnamefont
  {Nielsen}},\ }\href {\doibase 10.1016/j.nuclphysa.2011.04.001,
  10.1016/j.nuclphysa.2008.10.015} {\bibfield  {journal} {\bibinfo  {journal}
  {Nucl.Phys.}\ }\textbf {\bibinfo {volume} {A815}},\ \bibinfo {pages} {53}
  (\bibinfo {year} {2009})},\ \Eprint {http://arxiv.org/abs/0804.4817}
  {arXiv:0804.4817 [hep-ph]} \BibitemShut {NoStop}%
\bibitem [{\citenamefont {Qiao}(2008)}]{Qiao:2007ce}%
  \BibitemOpen
  \bibfield  {author} {\bibinfo {author} {\bibfnamefont {C.-F.}\ \bibnamefont
  {Qiao}},\ }\href {\doibase 10.1088/0954-3899/35/7/075008} {\bibfield
  {journal} {\bibinfo  {journal} {J.Phys.}\ }\textbf {\bibinfo {volume}
  {G35}},\ \bibinfo {pages} {075008} (\bibinfo {year} {2008})},\ \Eprint
  {http://arxiv.org/abs/0709.4066} {arXiv:0709.4066 [hep-ph]} \BibitemShut
  {NoStop}%
\bibitem [{\citenamefont {Ding}(2009)}]{Ding:2008gr}%
  \BibitemOpen
  \bibfield  {author} {\bibinfo {author} {\bibfnamefont {G.-J.}\ \bibnamefont
  {Ding}},\ }\href {\doibase 10.1103/PhysRevD.79.014001} {\bibfield  {journal}
  {\bibinfo  {journal} {Phys.Rev.}\ }\textbf {\bibinfo {volume} {D79}},\
  \bibinfo {pages} {014001} (\bibinfo {year} {2009})},\ \Eprint
  {http://arxiv.org/abs/0809.4818} {arXiv:0809.4818 [hep-ph]} \BibitemShut
  {NoStop}%
\bibitem [{\citenamefont {Chen}\ \emph {et~al.}(2011)\citenamefont {Chen},
  \citenamefont {He},\ and\ \citenamefont {Liu}}]{Chen:2010nv}%
  \BibitemOpen
  \bibfield  {author} {\bibinfo {author} {\bibfnamefont {D.-Y.}\ \bibnamefont
  {Chen}}, \bibinfo {author} {\bibfnamefont {J.}~\bibnamefont {He}}, \ and\
  \bibinfo {author} {\bibfnamefont {X.}~\bibnamefont {Liu}},\ }\href {\doibase
  10.1103/PhysRevD.83.054021} {\bibfield  {journal} {\bibinfo  {journal}
  {Phys.Rev.}\ }\textbf {\bibinfo {volume} {D83}},\ \bibinfo {pages} {054021}
  (\bibinfo {year} {2011})},\ \Eprint {http://arxiv.org/abs/1012.5362}
  {arXiv:1012.5362 [hep-ph]} \BibitemShut {NoStop}%
\bibitem [{\citenamefont {Brambilla}\ \emph {et~al.}(2011)\citenamefont
  {Brambilla} \emph {et~al.}}]{Brambilla:2010cs}%
  \BibitemOpen
  \bibfield  {author} {\bibinfo {author} {\bibfnamefont {N.}~\bibnamefont
  {Brambilla}} \emph {et~al.},\ }\href {\doibase
  10.1140/epjc/s10052-010-1534-9} {\bibfield  {journal} {\bibinfo  {journal}
  {Eur. Phys. J.}\ }\textbf {\bibinfo {volume} {C71}},\ \bibinfo {pages} {1534}
  (\bibinfo {year} {2011})},\ \Eprint {http://arxiv.org/abs/1010.5827}
  {arXiv:1010.5827 [hep-ph]} \BibitemShut {NoStop}%
\bibitem [{\citenamefont {Close}\ and\ \citenamefont
  {Swanson}(2005{\natexlab{b}})}]{PhysRevD.72.094004}%
  \BibitemOpen
  \bibfield  {author} {\bibinfo {author} {\bibfnamefont {F.~E.}\ \bibnamefont
  {Close}}\ and\ \bibinfo {author} {\bibfnamefont {E.~S.}\ \bibnamefont
  {Swanson}},\ }\href {\doibase 10.1103/PhysRevD.72.094004} {\bibfield
  {journal} {\bibinfo  {journal} {Phys. Rev. D}\ }\textbf {\bibinfo {volume}
  {72}},\ \bibinfo {pages} {094004} (\bibinfo {year}
  {2005}{\natexlab{b}})}\BibitemShut {NoStop}%
\bibitem [{\citenamefont {Danilkin}\ and\ \citenamefont
  {Simonov}(2010)}]{Danilkin:2010cc}%
  \BibitemOpen
  \bibfield  {author} {\bibinfo {author} {\bibfnamefont {I.}~\bibnamefont
  {Danilkin}}\ and\ \bibinfo {author} {\bibfnamefont {Y.}~\bibnamefont
  {Simonov}},\ }\href {\doibase 10.1103/PhysRevLett.105.102002} {\bibfield
  {journal} {\bibinfo  {journal} {Phys.Rev.Lett.}\ }\textbf {\bibinfo {volume}
  {105}},\ \bibinfo {pages} {102002} (\bibinfo {year} {2010})},\ \Eprint
  {http://arxiv.org/abs/1006.0211} {arXiv:1006.0211 [hep-ph]} \BibitemShut
  {NoStop}%
\bibitem [{\citenamefont {Coito}\ \emph {et~al.}(2011)\citenamefont {Coito},
  \citenamefont {Rupp},\ and\ \citenamefont {van Beveren}}]{Coito:2010if}%
  \BibitemOpen
  \bibfield  {author} {\bibinfo {author} {\bibfnamefont {S.}~\bibnamefont
  {Coito}}, \bibinfo {author} {\bibfnamefont {G.}~\bibnamefont {Rupp}}, \ and\
  \bibinfo {author} {\bibfnamefont {E.}~\bibnamefont {van Beveren}},\ }\href
  {\doibase 10.1140/epjc/s10052-011-1762-7} {\bibfield  {journal} {\bibinfo
  {journal} {Eur.Phys.J.}\ }\textbf {\bibinfo {volume} {C71}},\ \bibinfo
  {pages} {1762} (\bibinfo {year} {2011})},\ \Eprint
  {http://arxiv.org/abs/1008.5100} {arXiv:1008.5100 [hep-ph]} \BibitemShut
  {NoStop}%
\bibitem [{\citenamefont {Zhang}\ \emph {et~al.}(2009)\citenamefont {Zhang},
  \citenamefont {Meng},\ and\ \citenamefont {Zheng}}]{Zhang:2009bv}%
  \BibitemOpen
  \bibfield  {author} {\bibinfo {author} {\bibfnamefont {O.}~\bibnamefont
  {Zhang}}, \bibinfo {author} {\bibfnamefont {C.}~\bibnamefont {Meng}}, \ and\
  \bibinfo {author} {\bibfnamefont {H.}~\bibnamefont {Zheng}},\ }\href
  {\doibase 10.1016/j.physletb.2009.09.033} {\bibfield  {journal} {\bibinfo
  {journal} {Phys.Lett.}\ }\textbf {\bibinfo {volume} {B680}},\ \bibinfo
  {pages} {453} (\bibinfo {year} {2009})},\ \Eprint
  {http://arxiv.org/abs/0901.1553} {arXiv:0901.1553 [hep-ph]} \BibitemShut
  {NoStop}%
\bibitem [{\citenamefont {Abe}\ \emph {et~al.}(2007)\citenamefont {Abe} \emph
  {et~al.}}]{Abe:2007jna}%
  \BibitemOpen
  \bibfield  {author} {\bibinfo {author} {\bibfnamefont {K.}~\bibnamefont
  {Abe}} \emph {et~al.} (\bibinfo {collaboration} {Belle Collaboration}),\
  }\href {\doibase 10.1103/PhysRevLett.98.082001} {\bibfield  {journal}
  {\bibinfo  {journal} {Phys.Rev.Lett.}\ }\textbf {\bibinfo {volume} {98}},\
  \bibinfo {pages} {082001} (\bibinfo {year} {2007})},\ \Eprint
  {http://arxiv.org/abs/hep-ex/0507019} {arXiv:hep-ex/0507019 [hep-ex]}
  \BibitemShut {NoStop}%
\bibitem [{\citenamefont {Pakhlov}\ \emph {et~al.}(2008)\citenamefont {Pakhlov}
  \emph {et~al.}}]{Abe:2007sya}%
  \BibitemOpen
  \bibfield  {author} {\bibinfo {author} {\bibfnamefont {P.}~\bibnamefont
  {Pakhlov}} \emph {et~al.} (\bibinfo {collaboration} {Belle Collaboration}),\
  }\href {\doibase 10.1103/PhysRevLett.100.202001} {\bibfield  {journal}
  {\bibinfo  {journal} {Phys.Rev.Lett.}\ }\textbf {\bibinfo {volume} {100}},\
  \bibinfo {pages} {202001} (\bibinfo {year} {2008})},\ \Eprint
  {http://arxiv.org/abs/0708.3812} {arXiv:0708.3812 [hep-ex]} \BibitemShut
  {NoStop}%
\bibitem [{\citenamefont {Aubert}\ \emph {et~al.}(2007)\citenamefont {Aubert}
  \emph {et~al.}}]{Aubert:2006mi}%
  \BibitemOpen
  \bibfield  {author} {\bibinfo {author} {\bibfnamefont {B.}~\bibnamefont
  {Aubert}} \emph {et~al.} (\bibinfo {collaboration} {BaBar Collaboration}),\
  }\href {\doibase 10.1103/PhysRevD.76.111105} {\bibfield  {journal} {\bibinfo
  {journal} {Phys.Rev.}\ }\textbf {\bibinfo {volume} {D76}},\ \bibinfo {pages}
  {111105} (\bibinfo {year} {2007})},\ \Eprint
  {http://arxiv.org/abs/hep-ex/0607083} {arXiv:hep-ex/0607083 [hep-ex]}
  \BibitemShut {NoStop}%
\bibitem [{\citenamefont {Chao}(2008)}]{Chao:2007it}%
  \BibitemOpen
  \bibfield  {author} {\bibinfo {author} {\bibfnamefont {K.-T.}\ \bibnamefont
  {Chao}},\ }\href {\doibase 10.1016/j.physletb.2008.02.039} {\bibfield
  {journal} {\bibinfo  {journal} {Phys.Lett.}\ }\textbf {\bibinfo {volume}
  {B661}},\ \bibinfo {pages} {348} (\bibinfo {year} {2008})},\ \Eprint
  {http://arxiv.org/abs/0707.3982} {arXiv:0707.3982 [hep-ph]} \BibitemShut
  {NoStop}%
\bibitem [{\citenamefont {Guo}\ and\ \citenamefont
  {Meissner}(2012)}]{Guo:2012tv}%
  \BibitemOpen
  \bibfield  {author} {\bibinfo {author} {\bibfnamefont {F.-K.}\ \bibnamefont
  {Guo}}\ and\ \bibinfo {author} {\bibfnamefont {U.-G.}\ \bibnamefont
  {Meissner}},\ }\href {\doibase 10.1103/PhysRevD.86.091501} {\bibfield
  {journal} {\bibinfo  {journal} {Phys.Rev.}\ }\textbf {\bibinfo {volume}
  {D86}},\ \bibinfo {pages} {091501} (\bibinfo {year} {2012})},\ \Eprint
  {http://arxiv.org/abs/1208.1134} {arXiv:1208.1134 [hep-ph]} \BibitemShut
  {NoStop}%
\bibitem [{\citenamefont {Uehara}\ \emph {et~al.}(2006)\citenamefont {Uehara}
  \emph {et~al.}}]{Uehara:2005qd}%
  \BibitemOpen
  \bibfield  {author} {\bibinfo {author} {\bibfnamefont {S.}~\bibnamefont
  {Uehara}} \emph {et~al.} (\bibinfo {collaboration} {Belle Collaboration}),\
  }\href {\doibase 10.1103/PhysRevLett.96.082003} {\bibfield  {journal}
  {\bibinfo  {journal} {Phys.Rev.Lett.}\ }\textbf {\bibinfo {volume} {96}},\
  \bibinfo {pages} {082003} (\bibinfo {year} {2006})},\ \Eprint
  {http://arxiv.org/abs/hep-ex/0512035} {arXiv:hep-ex/0512035 [hep-ex]}
  \BibitemShut {NoStop}%
\bibitem [{\citenamefont {Aubert}\ \emph {et~al.}(2010)\citenamefont {Aubert}
  \emph {et~al.}}]{Aubert:2010ab}%
  \BibitemOpen
  \bibfield  {author} {\bibinfo {author} {\bibfnamefont {B.}~\bibnamefont
  {Aubert}} \emph {et~al.} (\bibinfo {collaboration} {BaBar Collaboration}),\
  }\href {\doibase 10.1103/PhysRevD.81.092003} {\bibfield  {journal} {\bibinfo
  {journal} {Phys.Rev.}\ }\textbf {\bibinfo {volume} {D81}},\ \bibinfo {pages}
  {092003} (\bibinfo {year} {2010})},\ \Eprint {http://arxiv.org/abs/1002.0281}
  {arXiv:1002.0281 [hep-ex]} \BibitemShut {NoStop}%
\end{thebibliography}%

\end{document}